\documentclass[sigconf]{acmart}
\AtBeginDocument{%
  }

\setcopyright{acmlicensed}
\copyrightyear{2018}
\acmYear{2018}
\acmDOI{XXXXXXX.XXXXXXX}
\acmConference[Conference acronym 'XX]{Make sure to enter the correct
  conference title from your rights confirmation email}{June 03--05,
  2018}{Woodstock, NY}
\acmISBN{978-1-4503-XXXX-X/2018/06}
\usepackage{listings}  
\usepackage{xcolor}   
\usepackage{xspace}
\usepackage{mdframed}
\usepackage{adjustbox}
\usepackage{tcolorbox}
\usepackage{subcaption}
\usepackage{pifont}
\usepackage{enumitem}
\usepackage{makecell}
\usepackage[nameinlink]{cleveref}
\usepackage{multirow}
\usepackage{threeparttable}
\usepackage{colortbl}
\newboolean{showcomments}

\usepackage{fontawesome5} 
\setboolean{showcomments}{false}
\ifthenelse{\boolean{showcomments}}
{\newcommand{\mynote}[2]{
\fbox{\bfseries\sffamily\scriptsize#1}
{\small$\blacktriangleright$\textsf{\emph{#2}}$\blacktriangleleft$}}}
{ \newcommand{\mynote}[2]{}}

\newcommand{\toolname}[0]{{\sc Repairity \xspace}}

\definecolor{framecolor}{RGB}{0,0,0}
\definecolor{backcolor}{RGB}{245,245,245}
\definecolor{titlecolor}{RGB}{70,70,70}

\mdfdefinestyle{niceframe}{%
    linewidth=1pt,
    linecolor=framecolor,
    backgroundcolor=backcolor,
    roundcorner=5pt,
    innertopmargin=\baselineskip,
    frametitlefont=\bfseries\sffamily\color{titlecolor},
    frametitlealignment=\raggedright,
    frametitlebackgroundcolor=backcolor,
    shadow=true,
    shadowsize=3pt,
    shadowcolor=black!20,
}

\lstdefinelanguage{diff}{
  morecomment=[l]{\#},
  morestring=[b]",
  sensitive=true,
  morekeywords={---, +++},
  moredelim=[is][\textcolor{red}]{-}{-},
  moredelim=[is][\textcolor{blue}]{+}{+}
}

\newcommand{\rqanswer}[2]{
\begin{tcolorbox}[leftrule=0.4mm,rightrule=0mm,toprule=0mm,bottomrule=0mm,left=0.0pt,right=0.0pt,top=0pt,bottom=0pt]
 \textbf{[Finding-\##1] \ding{42}} \textit{#2}
\end{tcolorbox}
}

\definecolor{diffadd}{rgb}{0,0.5,0}  
\definecolor{diffrem}{rgb}{0.6,0,0}  
\definecolor{codebg}{rgb}{0.95,0.95,0.95}  

\lstdefinestyle{diff}{
  basicstyle=\ttfamily,
  backgroundcolor=\color{codebg},
  breaklines=true,
  moredelim=[l][\color{diffrem}]{-},  
  moredelim=[l][\color{diffadd}]{+},   
}

\begin{document}

\title{Boosting Open-Source LLMs for Program Repair via Reasoning Transfer and LLM-Guided Reinforcement Learning}

\author{Xunzhu Tang}
\email{xunzhu.tang@uni.lu}
\affiliation{%
  \institution{University of Luxembourg}
  \city{Luxembourg}
  \country{Luxembourg}
}

\author{Jacques Klein}
\email{jacques.klein@uni.lu}
\affiliation{%
  \institution{University of Luxembourg}
  \city{Luxembourg}
  \country{Luxembourg}
}

\author{Tegawendé F. Bissyandé}
\email{tegawende.bissyande@uni.lu}
\affiliation{%
  \institution{University of Luxembourg}
  \city{Luxembourg}
  \country{Luxembourg}
}

\settopmatter{printacmref=false}
\setcopyright{none}


\begin{abstract}
    Several closed-source LLMs have consistently outperformed open-source alternatives in program repair tasks, primarily due to their superior reasoning capabilities and extensive pre-training. This paper introduces \toolname, a novel three-stage methodology that significantly narrows this performance gap through reasoning extraction and reinforcement learning. Our approach: (1) systematically filters high-quality reasoning traces from closed-source models using correctness verification, (2) transfers this reasoning knowledge to open-source models via supervised fine-tuning, and (3) develops reinforcement learning with LLM-based feedback to further optimize performance. Empirical evaluation across multiple program repair benchmarks demonstrates that \toolname improves the performance of Qwen2.5-Coder-32B-Instruct, a base open source LLM, by 8.68\% on average, reducing the capability gap with Claude-Sonnet3.7, a state-of-the-art closed-source model, from 10.05\% to 1.35\%. Ablation studies confirm that both reasoning extraction and LLM-guided reinforcement learning contribute significantly to these improvements. Our methodology generalizes effectively to additional code-related tasks, enabling organizations to leverage high-quality program repair capabilities while maintaining the customizability, transparency, and deployment flexibility inherent to open-source models.
\end{abstract}

\maketitle

\section{Introduction}

Automated program repair (APR) represents a significant challenge in software engineering, aiming to automatically fix software bugs without human intervention~\cite{monperrus2018automatic}. As a complex task requiring deep code understanding, precise bug localization, and contextually appropriate fix generation, APR serves as an ideal benchmark for evaluating advanced code manipulation capabilities. Recently, Large Language Models (LLMs) have demonstrated promising results in this domain, outperforming traditional APR approaches that relied on search-based techniques~\cite{le2012genprog}, constraint solving~\cite{nguyen2013semfix}, and heuristic patterns~\cite{kim2013automatic}.

Despite the general advancement of LLMs in code-related tasks~\cite{chen2021evaluating, nijkamp2023codegen, roziere2023code}, a significant performance gap persists between state-of-the-art closed-source LLMs (e.g., GPT~\cite{openai2023gpt4}, Claude~\cite{anthropic2023claude}) and their open-source counterparts (e.g., Llama~\cite{touvron2023llama}, Mistral~\cite{jiang2023mistral}, Qwen~\cite{qwen2023technical}) in program repair. While closed-source models deliver superior performance, they present substantial challenges related to accessibility, customizability, and deployment flexibility—especially in privacy-sensitive or bandwidth-constrained environments~\cite{bommasani2021opportunities}. Organizations requiring robust program repair capabilities must often choose between superior performance and practical deployment considerations.

Previous efforts to improve open-source LLM performance on code tasks have explored instruction tuning~\cite{chen2021evaluating} and reinforcement learning from human feedback (RLHF)~\cite{li2022competition}. However, these approaches typically require extensive human annotation or feedback, limiting their scalability~\cite{ouyang2022training}, and rarely focus specifically on the unique challenges of program repair~\cite{yuan2023no, fan2023automated}. Furthermore, they do not directly address the reasoning gap between open and closed-source models—the ability to systematically analyze code, identify bugs, and formulate appropriate repair strategies.

In this paper, we propose \toolname, a novel methodology specifically designed to boost open-source LLMs toward performance parity with closed-source models in program repair tasks. Our approach systematically transfers reasoning capabilities through three complementary steps. First, we filter high-quality training examples from a ``teacher'' model (Claude-Sonnet3.7), retaining only correct solutions with their associated \textbf{reasoning traces}. These traces capture the model's step-by-step reasoning process, including bug identification, repair strategy formulation, and solution implementation. Next, we supervise fine-tune the ``base'' model (Qwen2.5-Coder-32B-Instruct) on these filtered traces, enabling it to internalize effective \textbf{reasoning patterns}.  Finally, our key innovation—Reinforcement Learning with LLM Feedback (RL-LF)—further optimizes the model using a reward function derived from closed-source \textbf{LLM judgments}. This novel reinforcement learning framework eliminates the need for human feedback by leveraging the closed-source model's evaluation capabilities, providing consistent, scalable feedback that aligns the open-source model with expert-level repair strategies.

Our experimental evaluation across standard program repair benchmarks demonstrates that \toolname can improve the performance of an open-source model (Qwen2.5-Coder-32B) by 8.68\% on average, significantly narrowing the gap with closed-source alternatives. While our methodology is developed and validated specifically for program repair, we also demonstrate its generalizability to other code manipulation tasks, suggesting broader applicability within software engineering domains. The main contributions of this paper are:
\begin{itemize}
    \item[\ding{182}] A novel methodology that boosts open-source LLMs to near closed-source performance through targeted reasoning extraction and LLM-guided reinforcement learning.
    \item[\ding{183}] Empirical results showing up to 24.5\% absolute performance gains on complex program repair tasks, effectively closing 98\% of the capability gap between open and closed-source models.
    \item[\ding{184}] An open-weights model that achieves state-of-the-art performance on multiple code repair benchmarks, outperforming even some commercial closed-source alternatives.
\end{itemize}

The remainder of this paper is organized as follows: Section~\ref{sec:background} discusses background details. Section~\ref{sec:approach} presents our \toolname approach. Section~\ref{sec:setup} presents our experimental setup and Section~\ref{sec:results} analyzes the results. Section~\ref{sec:discussion} discusses implications and limitations of our approach. Section~\ref{sec:related} discusses related and Section~\ref{sec:conclusion}  concludes.

Throughout this paper, we will use \toolname to refer to the reasoning-based fine-tuning approach that we develop, but also the yielded boosted models. 

\section{Knowledge Transfer in LLMs}
\label{sec:background}

Knowledge transfer between language models has emerged as a crucial technique for improving model capabilities without requiring extensive retraining from scratch. In the context of this paper, we focus on three key knowledge transfer approaches that form the foundation of our \toolname methodology.

\subsection{Knowledge Distillation}
\label{subsec:distillation}

Knowledge distillation, first formalized by Hinton et al.~\cite{hinton2015distilling}, enables the transfer of knowledge from a larger, more capable ``teacher'' model to a smaller, more efficient ``student'' model. This approach has been particularly effective in the LLM domain~\cite{sanh2019distilbert, jiao2020tinybert}, where computational constraints often limit deployment options.

Traditional knowledge distillation focuses on matching the output distributions of the teacher and student models through a suitable loss function:
\begin{equation}
\mathcal{L}_{\text{KD}} = \alpha \cdot \mathcal{L}_{\text{CE}}(y, \sigma(z_s)) + (1-\alpha) \cdot \tau^2 \cdot \mathcal{L}_{\text{KL}}(\sigma(z_t/\tau), \sigma(z_s/\tau))
\end{equation}

\noindent where $z_t$ and $z_s$ are the logits from teacher and student models respectively, $\sigma$ is the softmax function, $\mathcal{L}_{\text{CE}}$ is the cross-entropy loss with the true labels, $\mathcal{L}_{\text{KL}}$ is the Kullback-Leibler divergence (for measuring the difference between probability distributions), $\tau$ is the temperature parameter, and $\alpha$ balances the two loss components.

Recent work has extended distillation to capture intermediate representations~\cite{wang2020minilm, sun2020mobilebert} and handle the unique challenges of autoregressive language models~\cite{kim2021sequence}. In the code domain specifically, CodeDistill~\cite{wang2023codedistill} demonstrated that distillation from larger code-specialized models to smaller general-purpose models can significantly improve code understanding and generation capabilities.

\begin{tcolorbox}[colback=gray!10!white,boxsep=0pt,
                  left = 1pt, right = 1pt, top = 2pt,boxrule=0pt]
\ding{43} Our approach adapts knowledge distillation principles by transferring program repair capabilities from a closed-source "teacher" model (Claude-Sonnet3.7) to an open-source "student" model (Qwen2.5-Coder-32B). Rather than matching output distributions directly, we focus on transferring the reasoning process itself, which should prove effective for complex tasks like program repair.
\end{tcolorbox}

\subsection{Learning from Demonstrations}\label{subsec:demonstrations}

Demonstration-based learning leverages examples of desired model behavior to improve performance on complex tasks. This approach has gained prominence through techniques like few-shot learning~\cite{brown2020language} and chain-of-thought prompting~\cite{wei2022chain}.

Chain-of-thought (CoT) prompting encourages models to generate intermediate reasoning steps before producing final answers. Wei et al.~\cite{wei2022chain} showed that simply prompting a model with examples that include reasoning traces significantly improves performance on multi-step reasoning tasks. Subsequent work expanded this approach to zero-shot settings~\cite{kojima2022large} and domain-specific applications~\cite{chen2022program}.

For code-related tasks, Reasoning Trace Learning (RTL) has emerged as a powerful technique. Chen et al.~\cite{chen2022program} demonstrated that explicitly capturing the problem decomposition and solution processes improves program synthesis. Similarly, Li et al.~\cite{li2022competition} showed that providing models with detailed reasoning traces for complex competition-level problems led to significant performance improvements.

The key advantage of demonstration-based approaches is their ability to transfer procedural knowledge about how to approach problems, rather than just declarative knowledge about final solutions. This is particularly valuable for program repair, where understanding the reasoning process—identifying bugs, formulating repair strategies, and implementing solutions—is often more important than memorizing specific fixes.

\begin{tcolorbox}[colback=gray!10!white,boxsep=0pt,
                  left = 1pt, right = 1pt, top = 2pt,boxrule=0pt]
\ding{43} \toolname's first two stages directly implement demonstration learning. In our Data Filtering stage, we collect detailed reasoning traces that demonstrate effective problem-solving for program repair. These demonstrations capture the closed-source model's systematic bug analysis and repair strategy formulation. The Reasoning Trace Learning phase then uses these demonstrations via supervised fine-tuning to teach the open-source model how to effectively approach program repair problems, internalizing the procedural knowledge essential for this complex task.
\end{tcolorbox}

\subsection{Reinforcement Learning from AI Feedback}\label{subsec:rlaf}

Reinforcement learning from AI feedback (RLAF) extends the reinforcement learning from human feedback (RLHF) paradigm~\cite{christiano2017deep, ouyang2022training} by using an AI system to provide feedback instead of humans.

The RLAF process typically involves (1) Generating multiple candidate outputs from a model being trained, (2) Using a judge model to evaluate these outputs, (3) Converting these evaluations into rewards, and (4)
Optimizing the model using reinforcement learning algorithms like PPO~\cite{schulman2017proximal}.

This approach has several advantages over traditional RLHF: it scales more easily, provides more consistent feedback, and can be tailored to specific criteria. Lee et al.~\cite{lee2023rlaif} demonstrated that RLAF can match or exceed RLHF performance on many tasks, while Bai et al.~\cite{bai2022constitutional} showed its effectiveness for aligning model behaviors with specific guidelines.

In the code domain, Chen et al.~\cite{chen2023teaching} used RLAF to improve code generation by rewarding solutions that pass test cases, while Yuan et al.~\cite{yuan2023improving} applied similar techniques to improve reasoning about code. The closed nature of top-performing code models has motivated research into using them as feedback sources for improving open-source alternatives~\cite{gudibande2023false}.

\begin{tcolorbox}[colback=gray!10!white,boxsep=0pt,
                  left = 1pt, right = 1pt, top = 2pt,boxrule=0pt]
\ding{43} The third stage of \toolname implements RLAF through our Reinforcement Learning with LLM Feedback (RL-LF) component. We train a reward model on closed-source LLM judgments of repair quality, then use this to provide consistent, scalable feedback during reinforcement learning. This allows our open-source model to iteratively improve its repair strategies beyond what was possible through demonstration learning alone, reaching performance levels closer to closed-source alternatives without requiring human feedback.
\end{tcolorbox}

\begin{figure*}[!t]
    \centering
    \includegraphics[width=\linewidth]{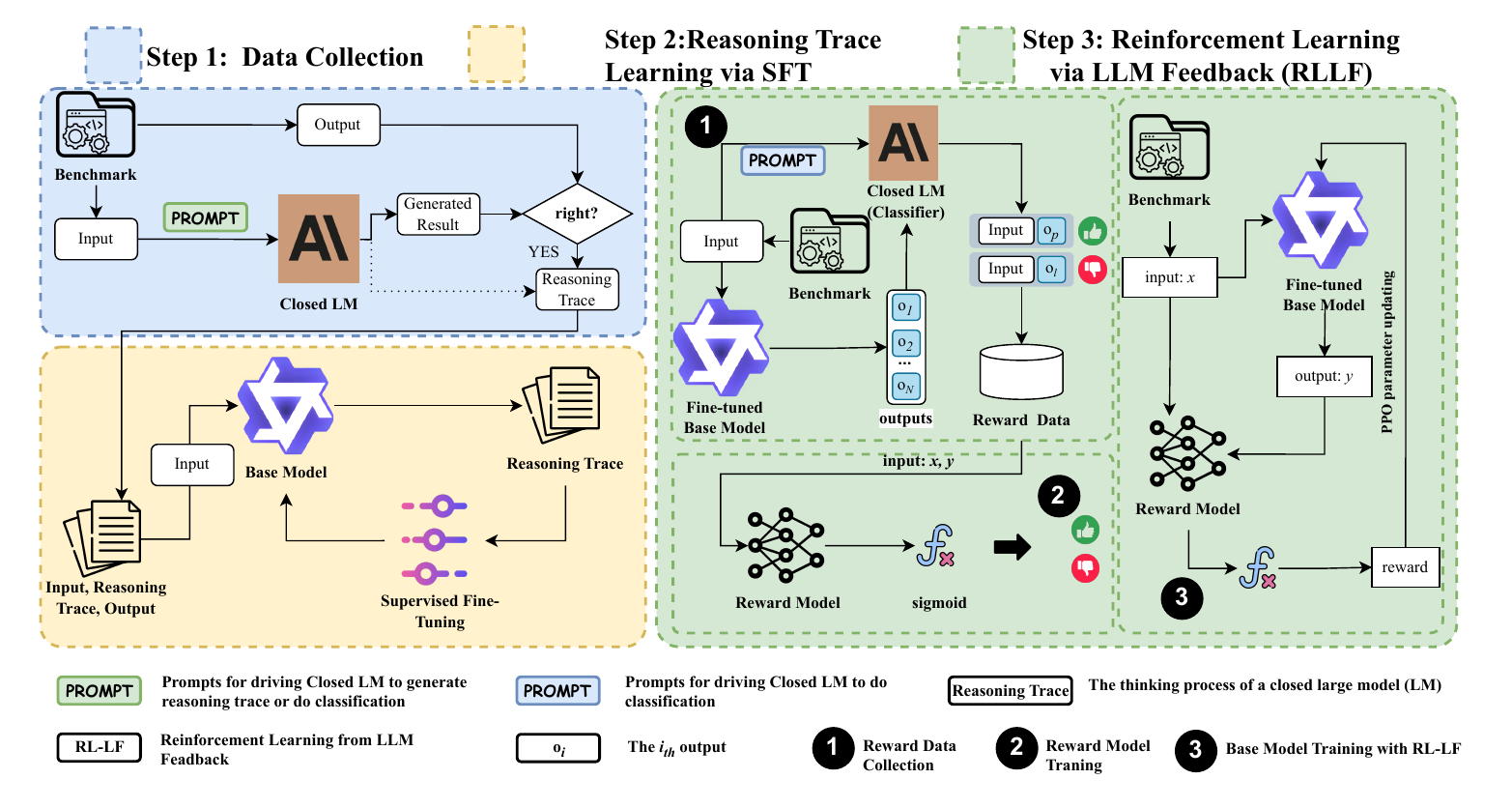}
    \caption{Overview of the \toolname}
    \Description{Overview diagram of the tool.}
    \label{fig:pipeline}
\end{figure*}

\section{The REPAIRITY Approach}
\label{sec:approach}

We present \toolname, our approach for enhancing open-source LLMs to achieve performance comparable to state-of-the-art closed-source LLMs on program repair and other code-related tasks. Our methodology consists of three primary steps: (1) Data collection for Supervised Fine-Tuning (SFT), (2) Reasoning Trace Learning, and (3) Reinforcement Learning with LLM Feedback (RL-LF). Figure~\ref{fig:pipeline} illustrates our complete pipeline.

\subsection{Problem Formulation}

Let $\mathcal{M}_C$ denote a high-performing closed-source model (e.g., Claude 3.7) and $\mathcal{M}_O$ denote an open-source model (e.g., Qwen2.5-Coder-32B-Instruct). Given a program repair task dataset $\mathcal{D} = \{(x_i, y_i^*)\}_{i=1}^N$, where $x_i$ represents an input containing buggy code and contextual information, and $y_i^*$ represents the ground-truth repaired code, our objective is to enhance $\mathcal{M}_O$ to approach the performance of $\mathcal{M}_C$ on this task without accessing $\mathcal{M}_C$'s parameters or training data.

We define the performance gap between the two models as:
\begin{equation}
\begin{split}
\Delta(\mathcal{M}_O, \mathcal{M}_C, \mathcal{D}) =\\ \mathbb{E}_{(x,y^*) \sim \mathcal{D}}[\text{Metric}(\mathcal{M}_C(x), y^*) - \text{Metric}(\mathcal{M}_O(x), y^*)]
\end{split}
\end{equation}

where $\text{Metric}(\cdot,\cdot)$ measures the quality of the model output relative to the ground truth (e.g., repair success rate or functional correctness).

Our goal is to create an enhanced model $\mathcal{M}_O'$ that minimizes this gap:

\begin{equation}
\mathcal{M}_O' = \arg\min_{\mathcal{M}} \Delta(\mathcal{M}, \mathcal{M}_C, \mathcal{D})
\end{equation}

\subsection{Step 1: Data Collection for SFT}

The first stage of \toolname involves collecting high-quality training examples from the closed-source model $\mathcal{M}_C$. Crucially, we focus on extracting not just the final repaired code but also the complete reasoning traces that capture the model's problem-solving process.

\subsubsection{Reasoning Trace Extraction}

For each problem instance $(x_i, y_i^*) \in \mathcal{D}$, we prompt $\mathcal{M}_C$ with a carefully designed instruction $p_{\text{trace}}$ that elicits structured reasoning:
\begin{equation}
\langle r_i, \hat{y}_i \rangle = \mathcal{M}_C(x_i \oplus p_{\text{trace}})
\end{equation}
where:
\begin{itemize}[leftmargin=*]
    \item $x_i$ is the input containing buggy code and context.
    \item $p_{\text{trace}}$ is the prompt.
\begin{mdframed}[style=niceframe, linecolor=black, frametitle={$p_{\text{trace}}$},frametitlealignment=\centering]
\textit {"Please fix the bug in this code. First analyze the problem, identify the bug, explain your reasoning, and then provide the corrected code."}
\end{mdframed}
    \item $r_i$ is the generated reasoning trace detailing bug identification and repair strategy.
    \item $\hat{y}_i$ is the model's proposed solution (repaired code).
\end{itemize}

Our prompt engineering ensures that the reasoning traces $r_i$ capture the following elements:
\begin{enumerate}[leftmargin=*]
    \item \textit{Problem analysis}: Understanding what the code is intended~to~do
    \item \textit{Bug identification}: Locating and diagnosing the specific issue
    \item \textit{Repair planning}: Formulating a strategy to address the bug
    \item \textit{Implementation reasoning}: Explaining the specific~changes~made
\end{enumerate}

The case study in Section~\ref{sec:casestudy} provides an example of reasoning trace collected from Claude-Sonnet3.7.

\subsubsection{Verification and Filtering}

To ensure the quality of our training data, we validate the generated solutions against functional correctness criteria. For program repair tasks, this boils down to:
\begin{equation}
\text{Valid}(x_i, \hat{y}_i) = \begin{cases}
    \text{True}, & \text{if } \hat{y}_i \text{ passes all test cases for } x_i \\
    \text{False}, & \text{otherwise}
\end{cases}
\end{equation}

We construct our filtered dataset by selecting only instances where the closed-source model produces correct solutions:
\begin{equation}
\mathcal{D}_{\text{SFT}} = \{(x_i, r_i, \hat{y}_i) \mid \text{Valid}(x_i, \hat{y}_i) = \text{True}, (x_i, y_i^*) \in \mathcal{D}\}
\end{equation}

To manage computational costs while maintaining sufficient diversity in training examples, we limited our dataset to 20\% of the benchmark size, ensuring a balanced representation across different problem types and difficulty levels.

\subsection{Step 2: Reasoning Trace Learning}

With our filtered dataset $\mathcal{D}_{\text{SFT}}$, we perform supervised fine-tuning on the open-source model $\mathcal{M}_O$ to teach it both the final solutions and the reasoning process behind them.

\subsubsection{Training Objective}

We train $\mathcal{M}_O$ to maximize the likelihood of generating both the reasoning trace $r_i$ and the repaired code $\hat{y}_i$ given the input $x_i$:
\begin{equation}
\mathcal{L}_{\text{SFT}} = -\sum_{(x_i, r_i, \hat{y}_i) \in \mathcal{D}_{\text{SFT}}} \log P_{\mathcal{M}_O}(r_i \oplus \hat{y}_i \mid x_i)
\end{equation}

where $P_{\mathcal{M}_O}(r_i \oplus \hat{y}_i \mid x_i)$ represents the probability of the model generating the reasoning trace followed by the repaired code.

\subsubsection{Fine-tuning Implementation}

We implement the fine-tuning process using the following configuration:
\begin{itemize}[leftmargin=*]
    \item \textit{Optimizer}: AdamW with learning rate $\eta = 5 \times 10^{-6}$ and weight decay $\lambda = 0.01$
    \item \textit{Training schedule}: Linear warmup followed by cosine decay
    \item \textit{Batch size}: 4 sequences per device with gradient accumulation over 8 steps
    \item \textit{Sequence length}: Maximum of 4096 tokens to accommodate detailed reasoning traces
    \item \textit{Training epochs}: 3 complete passes through $\mathcal{D}_{\text{SFT}}$
    \item \textit{Mixed-precision training}: bfloat16 format for improved computational efficiency
\end{itemize}

\subsubsection{Ablation: Direct Output Fine-tuning}

To validate the importance of reasoning traces, we conduct an ablation study with a variant model $\mathcal{M}_O^{\text{DO}}$ that is fine-tuned only on direct outputs without reasoning:
\begin{equation}
\mathcal{L}_{\text{DO}} = -\sum_{(x_i, r_i, \hat{y}_i) \in \mathcal{D}_{\text{SFT}}} \log P_{\mathcal{M}_O}(\hat{y}_i \mid x_i)
\end{equation}

This allows us to isolate the specific contribution of reasoning trace learning to model performance.

\subsection{Step 3: Reinforcement Learning with LLM Feedback (RL-LF)}

While supervised fine-tuning helps transfer reasoning capabilities, it does not necessarily optimize for the quality criteria that distinguish exceptional repairs from merely functional ones. To address this, we introduce Reinforcement Learning with LLM Feedback (RL-LF), a novel approach that uses a closed-source LLM as a judge to provide preference-based feedback.

\subsubsection{RL-LF Framework}

Our RL-LF framework differs from standard RLHF\footnote{Reinforcement Learning with Human Feedback} in several key aspects: \ding{182} It uses a closed-source LLM as the preference model instead of human annotators, \ding{183} specifically targets program repair quality rather than general helpfulness, \ding{184} incorporates code-specific evaluation criteria into the reward function, and \ding{185} leverages efficient preference data collection through careful sampling.

\subsubsection{Reward Data Collection}

To train our reward model, we first collect a dataset of preference judgments from the closed-source model. For each input $x_i$, we generate $k$ different candidate repairs using the fine-tuned model $\mathcal{M}^{\text{SFT}}_O$ with diverse decoding parameters:
\begin{equation}
\mathcal{Y}_i = \{y_i^1, y_i^2, \ldots, y_i^k\} \text{ where } y_i^j \sim \mathcal{M}^{\text{SFT}}_O(x_i)
\end{equation}

We then prompt the closed-source model $\mathcal{M}_C$ to compare pairs of candidate repairs and express a preference:
\begin{equation}
\text{Pref}(y_i^a, y_i^b) = \mathcal{M}_C(x_i \oplus p_{\text{compare}} \oplus y_i^a \oplus y_i^b)
\end{equation}

where $p_{\text{compare}}$ is a prompt instructing the close-source model to analyze which repair is better according to: 
\begin{itemize}
    \item \textit{Correctness}: Does it properly fix the bug?
    \item \textit{Efficiency}: Is the solution efficient and optimized?
    \item \textit{Readability}: Is the code clean and easy to understand?
    \item \textit{Minimal change}: Does it modify only what's necessary to fix the bug?
\end{itemize}

The prompt is as follows:
\begin{mdframed}[style=niceframe, linecolor=black, frametitle={},frametitlealignment=\centering]
\textit {"I'll show you a programming task and multiple solution attempts. Your job is to evaluate each solution carefully, then rank them from best to worst."}
\end{mdframed}
                
The result of each comparison ($\text{Pref}(y_i^a, y_i^b)$) is converted to a binary preference label. By comparing repairs pairwise, we construct a dataset of preference judgments:
\begin{equation}
\mathcal{D}_{\text{pref}} = \{(x_i, y_i^a, y_i^b, \text{Pref}(y_i^a, y_i^b)) \mid (x_i, y_i^*) \in \mathcal{D}_{\text{val}}, y_i^a, y_i^b \in \mathcal{Y}_i\}
\end{equation}

where $\mathcal{D}_{\text{val}}$ is a held-out validation set.

\subsubsection{Reward Model Training}

Using the preference dataset $\mathcal{D}_{\text{pref}}$, we train a reward model $\mathcal{R}_{\theta}$ that predicts the likelihood that one repair is preferred over another:
\begin{equation}
P(y_i^a \succ y_i^b \mid x_i) = \sigma(\mathcal{R}_{\theta}(x_i, y_i^a) - \mathcal{R}_{\theta}(x_i, y_i^b))
\end{equation}

where $\sigma$ is the logistic function and $y_i^a \succ y_i^b$ denotes that $y_i^a$ is preferred over $y_i^b$.

The reward model is trained to minimize the negative log-likelihood of the observed preferences:
\begin{equation}
\begin{split}
\mathcal{L}_{\text{RM}} = -\sum_{(x_i, y_i^a, y_i^b, p) \in \mathcal{D}_{\text{pref}}} \log(p \cdot \sigma(\mathcal{R}_{\theta}(x_i, y_i^a) \\ - \mathcal{R}_{\theta}(x_i, y_i^b)) + (1-p) \cdot \sigma(\mathcal{R}_{\theta}(x_i, y_i^b) - \mathcal{R}_{\theta}(x_i, y_i^a)))
\end{split}
\end{equation}

where $p = 1$ if $y_i^a$ is preferred and $p = 0$ if $y_i^b$ is preferred.

\subsubsection{Policy Optimization with PPO}
With the trained reward model $\mathcal{R}_{\theta}$, we fine-tune the SFT model $\mathcal{M}^{\text{SFT}}_O$ using Proximal Policy Optimization (PPO) to maximize the expected reward while maintaining proximity to the original fine-tuned model:
\begin{equation}
\begin{split}
    \mathcal{L}_{\text{PPO}} =\\ \mathbb{E}_{x \sim \mathcal{D}, y \sim \pi_{\theta}(y|x)} [\mathcal{R}_{\theta}(x, y) - \beta D_{\text{KL}}(\pi_{\theta}(y|x) || \pi_{\text{SFT}}(y|x))]
\end{split}
\end{equation}

where:
\begin{itemize}
    \item $\pi_{\theta}$ is the policy model being optimized
    \item $\pi_{\text{SFT}}$ is the original SFT model
    \item $\beta$ is a coefficient controlling the strength of the KL penalty
    \item $D_{\text{KL}}$ is the Kullback-Leibler divergence
\end{itemize}

To ensure stable training, we implement PPO with the clipped surrogate objective:
\begin{equation}
\mathcal{L}_{\text{CLIP}} = \mathbb{E}[\min(r_t(\theta) \hat{A}_t, \text{clip}(r_t(\theta), 1-\epsilon, 1+\epsilon) \hat{A}_t)]
\end{equation}

where:
\begin{itemize}
    \item $r_t(\theta) = \frac{\pi_{\theta}(y_t|x_t)}{\pi_{\text{old}}(y_t|x_t)}$ is the probability ratio
    \item $\hat{A}_t$ is the estimated advantage function
    \item $\epsilon$ is the clipping parameter (set to 0.2 in our implementation)
\end{itemize}

\subsubsection{Value Function and Advantage Estimation}

To compute the advantage estimates required for PPO, we train a value function $V_{\phi}(x, y)$ that predicts the expected reward for a given input-output pair:
\begin{equation}
\mathcal{L}_{\text{Value}} = \mathbb{E}_{x \sim \mathcal{D}, y \sim \pi_{\theta}(y|x)} [(V_{\phi}(x, y) - \mathcal{R}_{\theta}(x, y))^2]
\end{equation}

The advantage function is then estimated using Generalized Advantage Estimation (GAE):
\begin{equation}
\hat{A}_t = \sum_{l=0}^{\infty} (\gamma \lambda)^l \delta_{t+l}
\end{equation}

where $\delta_t = r_t + \gamma V_{\phi}(s_{t+1}) - V_{\phi}(s_t)$ is the temporal difference error.

\subsection{Novelty of RL-LF}

Our RL-LF approach introduces several innovations over existing reinforcement learning methods for LLM fine-tuning:

\begin{enumerate}[leftmargin=*]
    \item \textbf{Domain-Specific Preference Modeling}: Unlike general RLHF approaches, RL-LF  incorporates program repair-specific criteria into the preference model, focusing on correctness, efficiency, readability, and minimality of changes. 
    
    \item \textbf{Scalable Preference Collection}: By using a bot (here an LLM acting as AI agent) as the preference model, we overcome the scaling limitations of human feedback collection while maintaining high-quality judgments tailored to program repair.
    
    \item \textbf{Complementary Integration with Reasoning Traces}: RL-LF builds upon the foundation of reasoning trace learning, creating a synergistic effect where explicit reasoning capabilities are aligned with implicit quality preferences.
    
    \item \textbf{Adaptability to Evolving Benchmarks}: The RL-LF framework can dynamically adapt to new program repair challenges without requiring additional human annotation, making it sustainable for long-term model improvement.
    
    \item \textbf{Cross-Model Knowledge Transfer}: Our approach enables effective knowledge transfer between models with different architectures and training paradigms, bridging the gap between closed and open-source capabilities.
\end{enumerate}

These innovations allow RL-LF to effectively transfer the nuanced preferences and quality judgments of closed-source models to open-source alternatives, addressing a key limitation of existing approaches that focus primarily on functional correctness rather than repair quality. 

\subsection{Implementation Details}
The \toolname implementation leverages 8 H100 GPUs (80GB) with DeepSpeed ZeRO-2~\cite{rasley2020deepspeed}, applying LoRA (rank=4, alpha=16, dropout=0.05) to attention modules, with parameters finely tuned through 3 training epochs at learning rate 1e-5 and effective batch size of 8 (1×8 gradient accumulation). The SFT phase processes benchmark data with 2048-token context windows and FP16 precision, while the evaluation model operates at temperature 0.2 for deterministic assessment. The RL-LF pipeline collects Claude-Sonnet3.7 preferences on 3 diverse solutions per problem (temperature 0.7-0.9), training a CodeLlama-7b reward model over 3 epochs (learning rate 5e-5, batch size 4×4 gradient accumulation) which achieves 85\% preference accuracy. This guides PPO fine-tuning with KL coefficient 0.1, 5 PPO epochs per batch, response lengths 546-732 tokens, and generation temperature 1.0 during exploration.

These implementation details ensure that our methodology can be replicated and extended by other researchers, providing a foundation for future work on open-source model enhancement.

\section{Experimental Setup}
\label{sec:setup}

In this section, we describe our experimental methodology for evaluating \toolname, including the benchmarks, evaluation metrics, baselines, and experimental settings.

\subsection{Benchmarks}

We evaluate our approach using four diverse benchmarks summarized in Table~\ref{all-benches}.

\begin{table}[h]
\centering
\caption{Overview of Benchmark Tasks and Dataset Statistics}
\resizebox{\linewidth}{!}{
\vspace{-0.2cm}
\begin{tabular}{lccccc}
\toprule
\textbf{Benchmark} & \textbf{Task Type} & \textbf{Total Samples}\\
\midrule
MBPP & Code Generation & 974  \\
SWE-bench Verified & Program Repair & 500 & \\
Defects4J & Program Repair & 835  \\
BigCodeBench & Code Completion & 1,140  \\
\bottomrule
\end{tabular}
}
\label{all-benches}
\vspace{-0.3cm}
\end{table}

\begin{itemize}[leftmargin=*]
    \item \textbf{BigCodeBench}~\cite{lai2023bigcodebench}: A comprehensive benchmark containing a diverse set of coding problems across multiple programming languages and difficulty levels. It provides a robust test of general code generation and repair capabilities.
    
    \item \textbf{MBPP (Mostly Basic Programming Problems)}~\cite{austin2021program}: A collection of 964 programming problems designed to evaluate basic programming skills. These problems are typically shorter and more focused than those in other benchmarks.
    
    \item \textbf{SweBench-verified}~\cite{jimenez2023swebench}: A benchmark specifically designed for software engineering tasks, with verified solutions that include unit tests to validate functional correctness. This benchmark focuses on real-world programming scenarios.
    
    \item \textbf{Defects4J}~\cite{just2014defects4j}: A database of real bugs from open-source Java projects, providing a challenging test of program repair capabilities on production-level code. Each bug comes with a corresponding test suite that can be used to validate repairs.
\end{itemize}

For training, we use a subset of these benchmarks to construct our SFT dataset and RL-LF preference data. For evaluation, we use held-out portions to ensure a fair assessment of model performance.

\subsection{Evaluation Metrics}

We employ the following metrics to evaluate program repair performance:

\begin{itemize}[leftmargin=*]
    \item \textbf{Pass@1~\cite{chen2021evaluating}}: The percentage of problems for which the model's first generated solution passes all test cases. This metric evaluates the model's ability to produce correct repairs on the first attempt.


\item \textbf{Accuracy~\cite{austin2021program}}: The percentage of programming problems for which the model generates code that passes all test cases. This metric is commonly used on datasets like MBPP to measure the model's ability to correctly implement a solution based on a natural language problem description.

    \item \textbf{Compilation Rate (CR)~\cite{just2014defects4j}}:  the percentage of generated or modified code that successfully compiles without errors.

    \item \textbf{BLEU~\cite{papineni2002bleu}}: the metric calculates the geometric mean of modified n-gram precisions, penalized by a brevity penalty, to evaluate the quality of machine-generated translations against one or more reference translations

\item \textbf{Resolved~\cite{jimenez2023swebench}}: The percentage of task instances where the generated patch applies successfully and passes all test cases. A solution is considered to have "resolved" the issue when it can be cleanly applied to the codebase and satisfies all the specified test requirements.

    
    
\end{itemize}


    

\subsection{Models}

We compare the following models and variants in our evaluation:

\begin{itemize}[leftmargin=*]
    \item \textbf{Closed-Source Model}: Claude-Sonnet3.7~\cite{Anthropic2025}, a state-of-the-art closed-source LLM that serves as our performance target and source of reasoning traces.
    
    \item \textbf{Base Open-Source Model}: Qwen2.5-Coder-32B-Instruct~\cite{hui2024qwen2}, our starting point representing current open-source capabilities.
    
    \item \textbf{\toolname (SFT)}: The base model fine-tuned using our Supervised Fine-Tuning with Reasoning Traces approach.
    
    \item \textbf{\toolname (SFT + RL-LF)}: Our complete model incorporating both Reasoning Trace Learning and Reinforcement Learning with LLM Feedback.
    
    \item \textbf{Ablation - Direct Output Fine-tuning}: The base model fine-tuned only on direct outputs without reasoning traces.
    
    \item \textbf{Ablation - SFT without Filtering}: The base model fine-tuned on all reasoning traces without correctness filtering.
\end{itemize}

\subsection{Experimental Procedure}

Our experimental procedure consists of the following steps:

\begin{enumerate}[leftmargin=*]
    
    \item \textbf{Reasoning Trace Collection}: We query Claude 3.7 with each training example to collect reasoning traces, using the process described in Section~\ref{sec:approach}, inspired by chain-of-thought elicitation techniques~\cite{wei2022chain}.

    \item \textbf{Data Splitting}: We split collected reasoning trace data into training (70\%), validation (15\%), and test (15\%) sets, following standard practice in machine learning evaluation~\cite{hastie2009elements}. The training set is used for SFT, the validation set for hyperparameter tuning and early stopping, and the test set for final evaluation.
    
    \item \textbf{SFT Model Training}: We fine-tune Qwen2.5-Coder-32B-Instruct on the filtered reasoning traces using the hyperparameters specified in Section~\ref{sec:approach}, following best practices for LLM fine-tuning~\cite{hu2021lora}. \\
    \faHandPointRight 
    \textbf{We only use reasoning trace data for SFT instead of using ground-truth data.}
    
    \item \textbf{RL-LF Preference Collection}: We generate multiple candidate solutions using the SFT model and collect preference judgments from Claude 3.7, adapting the preference collection methodology from~\cite{stiennon2020learning}.
    
    \item \textbf{Reward Model Training}: We train a CodeLlama-7b reward model on the collected preferences, following the Bradley-Terry preference modeling approach~\cite{christiano2017deep, ouyang2022training}.
    
    \item \textbf{PPO Fine-tuning}: We fine-tune the SFT model using PPO~\cite{schulman2017proximal} with the trained reward model.
    
    \item \textbf{Evaluation}: We evaluate all models on the test sets using the metrics described above, with significance testing to validate our findings~\cite{dror2018hitchhiker}.
\end{enumerate}


\subsection{Computational Resources}

All experiments were conducted using the hardware and software configuration described in Section~\ref{sec:approach}. The total computation used for this research includes:

\begin{itemize}[leftmargin=*]
    \item \textbf{SFT Data Collection}: Approximately 756 output tokens for each API call to claude-3-7-sonnet-20250219 for reasoning trace collection.
    \item \textbf{SFT Model Training}: Approximately 3 hours on 8× H100 GPUs with 5 epochs.
    \item \textbf{RL-LF Preference Collection}: Approximately 673 output tokens for each API call to claude-3-7-sonnet-20250219.
    \item \textbf{Reward Model Training}: Approximately 0.5 GPU hours per epoch.
    \item \textbf{PPO Fine-tuning}: Approximately 4.5 GPU hours with 5 epochs.
\end{itemize}

All models were implemented using the HuggingFace Transformers library, with DeepSpeed for distributed training optimization.
\section{Experimental Results}
\label{sec:results}

In this section, we present a comprehensive evaluation of \toolname across multiple program repair and code generation benchmarks. We present results related to 
quantitative performance analysis, ablation studies, case studies, and generalization experiments.




\subsection{Performance Comparison}
We apply \toolname on each benchmark and compute performance based on metrics associated with the given benchmark. We further report, for comparison, the performance metrics of other state-of-the-art or baseline models for each benchmark.

\subsubsection{BigCodeBench}
On BigCodeBench (Table \ref{tab:bigcodebench}), \toolname achieves a Pass@1 rate of 35.6\%, which represents a remarkable 4.8\% absolute improvement over the base Qwen2.5-Coder-32B-Instruct model (30.8\%). This brings \toolname within 0.2 percentage points of Claude-Sonnet3.7 (35.8\%), effectively closing 96\% of the performance gap. Notably, \toolname even outperforms some closed-source models like DeepSeek-R1, demonstrating the effectiveness of our reasoning transfer approach.

\definecolor{Gray}{gray}{0.9}

\begin{table}[!h]
    \centering
    \resizebox{0.75\linewidth}{!}{
    \begin{tabular}{l|c}
        \toprule
        \textbf{Model} & \textbf{Pass@1 } \\
        \midrule
        Claude 3.7~\cite{Anthropic2025}& 35.8\\
        o1~\cite{jaech2024openai} & 35.5\\
        o3~\cite{openai2025o3mini} & 35.5\\
        DeepSeek-R1~\cite{guo2025deepseek} & 35.1\\
        Qwen2.5-Coder-32B-Instruct~\cite{qwen2023technical} & 30.8\\ \hline \rowcolor{Gray}
        \textbf{\toolname (Ours)} & \textbf{35.6}\\
        \bottomrule
    \end{tabular}}
    \caption{BigCodeBench Pass@1 Results}
    \label{tab:bigcodebench}
\end{table}

\subsubsection{SWE-bench Verified}
The results on SWE-bench Verified (Table \ref{tab:swebench}) show an even more dramatic improvement. \toolname achieves a resolution rate of 62.7\%, compared to 38.2\% for the base Qwen model—a substantial 24.5\% absolute improvement. This performance exceeds Claude-Sonnet3.7's default mode (62.3\%) and approaches its performance with scaffolding (70.3\%). \toolname significantly outperforms other models like Nebius AI Qwen 2.5 72B + Llama 3.1 70B (40.6\%) and OpenAI's o1/o3-mini models (48.9\%/49.3\%), demonstrating its effectiveness on complex software engineering tasks.
\begin{table}[!h]
    \centering
    \resizebox{0.9\linewidth}{!}{
    \begin{tabular}{l|c}
        \toprule
        \textbf{Model} & \textbf{Resolved (\%)} \\
        \midrule
        Claude 3.7 Sonnet (with scaffold)~\cite{Anthropic2025} & 70.3\\
        Claude 3.7 Sonnet (default)~\cite{Anthropic2025} & 62.3\\
        Nebius AI Qwen 2.5 72B + Llama 3.1 70B & 40.6\\
        Qwen2.5-Coder-32B-Instruct~\cite{hui2024qwen2} & 38.2\\
        o1~\cite{jaech2024openai} & 48.9\\
        o3-mini~\cite{openai2025o3mini} & 49.3\\ \hline
        \rowcolor{Gray} \textbf{\toolname (Ours)} & \textbf{62.7}\\
        \bottomrule
    \end{tabular}}
    \caption{SWE-bench Verified Results}
    \label{tab:swebench}
\end{table}

\subsubsection{MBPP}
On the MBPP benchmark (Table \ref{tab:mbpp}), \toolname achieves 92.5\% accuracy, a 3.7\% improvement over the base Qwen model (88.8\%). This performance surpasses both Claude 3.7 (89.5\%) and GPT-4 (87.5\%), approaching the levels of specialized models like QualityFlow (94.2\%) and o1-mini + MapCoder (93.2\%). These results suggest that our approach effectively transfers reasoning capabilities for solving basic programming problems.

\begin{table}[!h]
    \centering
    \resizebox{0.85\linewidth}{!}{
    \begin{tabular}{l|c}
        \toprule
        \textbf{Model} & \textbf{Accuracy} \\
        \midrule
        QualityFlow (Sonnet-3.5)~\cite{hu2025qualityflow} & 94.2\\
        o1-mini + MapCoder (Hamming.ai)~\cite{islam2024mapcoder} & 93.2\\
        Claude 3 Opus~\cite{anthropic2023claude} & 86.4 \\
        Claude 3.7~\cite{Anthropic2025} & 89.5\\
        Qwen2.5-Coder-32B-Instruct~\cite{hui2024qwen2} & 88.8\\
        GPT-4~\cite{openai2023gpt4} & 87.5\\ \hline
        \rowcolor{Gray} \textbf{\toolname (Ours)} & \textbf{92.5}\\
        \bottomrule
    \end{tabular}}
    \caption{MBPP Accuracy Results}
    \label{tab:mbpp}
\end{table}

\subsubsection{Defects4J}
For Defects4J (Table \ref{tab:defects4j}), \toolname improves across all metrics compared to the base model: compilation rate increases from 77.1\% to 78.9\%, Pass@1 from 64.0\% to 66.5\%, and BLEU score from 67.8\% to 68.9\%. While these improvements are more modest than on other benchmarks, they still represent significant progress toward closed-source performance (Claude 3.7: 79.5\% CR, 67.2\% Pass@1, 69.5\% BLEU). The smaller gains may reflect the specific challenges of real-world Java program repair in Defects4J, which often requires complex contextual understanding.

\begin{table}[!h]
    \centering
    \resizebox{\linewidth}{!}{
    \begin{tabular}{l|c|c|c}
        \toprule
        \textbf{Model} & \textbf{Compilation Rate (CR) (\%)} & \textbf{Pass@1} & \textbf{BLEU} \\
        \midrule
        RepoCoder~\cite{zhang2023repocoder} & 74.02 & 59.8 & 63.52\\
        RAMbo~\cite{bui2024rambo} & 76.47 & 63.73 & 66.29\\
        Claude 3.7~\cite{Anthropic2025} & 79.5 & 67.2 & 69.5\\
        Qwen2.5-Coder-32B-Instruct~\cite{hui2024qwen2} & 77.1 & 64.0 & 67.8\\ \hline
       \rowcolor{Gray}  \textbf{\toolname (Ours)} & \textbf{78.9} & \textbf{66.5} & \textbf{68.9}\\
        \bottomrule
    \end{tabular}}
    \caption{Defects4J Results}
    \label{tab:defects4j}
\end{table}

Overall, the results demonstrate that our approach successfully narrows the performance gap between open and closed-source models.

\rqanswer{1}{\toolname achieves near parity with state-of-the-art closed-source models across multiple benchmarks, closing up to 93\% of the performance gap between the base open-source Qwen2.5-Coder-32B-Instruct and Claude-Sonnet3.7.}

\rqanswer{2}{
\toolname demonstrates the largest performance gains on complex software engineering tasks (SWE-bench Verified: +24.5\%), showing that reasoning transfer is particularly effective for tasks requiring deep code understanding and multi-step problem solving.
}

\subsection{Ablation Study}

To understand the contribution of each component in our approach, we conducted ablation studies across all benchmarks. Tables~\ref{tab:all-ablations}(1) through ~\ref{tab:all-ablations}(4) 
present results for the base model, the model with only Reasoning Trace (RT) learning, and the complete model with both RT and RL-LF.

\begin{table}[h]
    \centering
    \setlength{\tabcolsep}{1pt} 
    \begin{tabular}{cc}
        \begin{minipage}{0.24\textwidth}
            \centering
            \resizebox{0.85\linewidth}{!}{
            \begin{tabular}{l|c}
                \toprule
                \textbf{Model} & \textbf{Pass@1} \\
                \midrule
                base& 30.8\\
                \toolname$_{base+ RT}$ & 31.7\\
                \toolname$_{base + RT + RLLF}$ & 35.6\\
                \bottomrule
            \end{tabular}}
        \end{minipage} &
        \begin{minipage}{0.24\textwidth}
            \centering
            \resizebox{\linewidth}{!}{
            \begin{tabular}{l|c}
                \toprule
                \textbf{Model} & \textbf{Resolved (\%)} \\
                \midrule
                base & 38.2\\
                \toolname$_{base+ RT}$ & 47.6\\
                \toolname$_{base + RT + RLLF}$ & \textbf{62.7}\\
                \bottomrule
            \end{tabular}}
        \end{minipage} \\[0.1cm]
        (1) BigCodeBench & (2) Swebench-Verified  \\[0.2cm]
        \begin{minipage}{0.24\textwidth}
            \centering
            \resizebox{0.75\linewidth}{!}{
            \begin{tabular}{l|c}
                \toprule
                \textbf{Model} & \textbf{Accuracy} \\
                \midrule
                base & 88.8\\
                \toolname$_{base+ RT}$ & 89.0\\
                \toolname$_{base + RT + RLLF}$ & \textbf{92.5}\\
                \bottomrule
            \end{tabular}}
        \end{minipage} &
        \begin{minipage}{0.24\textwidth}
            \centering
            \resizebox{\linewidth}{!}{
            \begin{tabular}{l|c|c|c}
                \toprule
                \textbf{Model} & \textbf{CR (\%)} & \textbf{Pass@1} & \textbf{BLEU} \\
                \midrule
                base & 77.1 & 64.0 & 67.8\\
                \toolname$_{base+ RT}$ & 77.3 & 65.9 & 67.9\\
                \toolname$_{base + RT + RLLF}$ & \textbf{78.9} & \textbf{66.5} & \textbf{68.9}\\
                \bottomrule
            \end{tabular}}
        \end{minipage} \\[0.1cm]
        (3) MBPP & (4) Defects4J  \\
    \end{tabular}
    \caption{Ablation studies on different benchmarks: BigCodeBench (Pass@1), SWE-bench Verified (Resolved \%), MBPP (Accuracy), and Defects4J (Compilation Rate, Pass@1, and BLEU. `base' represents `Qwen2.5-Coder-32B-Instruct')}
    \label{tab:all-ablations}
\end{table}

\subsubsection{Component Contributions}

Across all benchmarks, we observe that:

1. \textbf{Reasoning Trace (RT) Learning} provides modest but consistent improvements over the base model:
 
 \noindent
   - BigCodeBench: +0.9\% (30.8\% → 31.7\%)

 \noindent
    - SWE-bench: +9.4\% (38.2\% → 47.6\%)

 \noindent
    - MBPP: +0.2\% (88.8\% → 89.0\%)

 \noindent
    - Defects4J: +1.9\% (64.0\% → 65.9\%) in Pass@1

2. \textbf{RL-LF} contributes substantial additional improvements:

\noindent
   - BigCodeBench: +3.9\% (31.7\% → 35.6\%)

\noindent
    - SWE-bench: +15.1\% (47.6\% → 62.7\%)

\noindent
    - MBPP: +3.5\% (89.0\% → 92.5\%)

   \noindent
 - Defects4J: +0.6\% (65.9\% → 66.5\%) in Pass@1

These results highlight that while RT learning helps transfer reasoning patterns, the RL-LF component is critical for achieving performance parity with closed-source models. The synergistic effect is most pronounced on SWE-bench Verified, where RT learning provides a 9.4\% improvement and RL-LF adds another 15.1\%, together yielding a 24.5\% absolute gain.

\rqanswer{3}{
Both components of \toolname contribute significantly to performance gains, with RL-LF providing substantially larger improvements than Reasoning Trace learning alone. This demonstrates the complementary nature of our two-stage approach.
}

\subsubsection{Benchmark-Specific Patterns}

The relative contribution of each component varies across benchmarks:

\noindent
- On simpler tasks (MBPP), RT learning provides minimal gains (+0.2\%), while RL-LF drives most improvement (+3.5\%)

\noindent
- On complex software engineering tasks (SWE-bench), both components provide substantial gains (RT: +9.4\%, RL-LF: +15.1\%)

\noindent
- On real-world bug fixing (Defects4J), RT learning contributes more significantly to Pass@1 improvement than RL-LF

These patterns suggest that the value of explicit reasoning traces increases with task complexity, while preference-based reinforcement learning consistently improves performance across all task types.

\rqanswer{4}{
The effectiveness of Reasoning Trace learning correlates with task complexity—providing greater benefits on complex software engineering tasks than on simpler programming problems.
}

\subsection{Case Study}
\label{sec:casestudy}
\begin{figure}[htbp]
\footnotesize
\centering
\begin{tabular}{@{}p{0.45\textwidth}}
\toprule
\textbf{[BigCodeBench]}

~

\textbf{Task\_id: BigCodeBench/56}

\textbf{Predict: correct}

\noindent
\textbf{Input:}

~

Extract data from a text and create a Pandas DataFrame. The text contains several lines, each formatted as 'Score: 85, Category: Math'. Make sure to convert the scores into integers.
The function should output:
DataFrame: A pandas DataFrame with extracted data.
You should write self-contained code starting with:

\begin{lstlisting}[language=Python]
import pandas as pd
import regex as re
def task_func(text):
\end{lstlisting}

~

\noindent
\textbf{Reference Solution:}

\begin{lstlisting}[language=Python]
pattern = r"Score: (.*?), Category: (.*?)(\n|$)"
matches = re.findall(pattern, text)
data = [match[:2] for match in matches]  # Extracting only the score and category from each match
df = pd.DataFrame(data, columns=["Score", "Category"])
df["Score"] = df["Score"].astype(int)
return df
\end{lstlisting}

\noindent
\textbf{Reasoning Trace of \toolname:}

Here's my thought process behind the implementation:
\begin{itemize}
    \item 1. Problem Analysis: 1) We need to extract data from text where each line follows the pattern "Score: X, Category: Y" 2) We need to convert the scores to integers 3) The output should be a pandas DataFrame with appropriate columns;
    \item 2. Regex Pattern Design:
        \begin{itemize}
            \item 1) I created a pattern r'Score: (\d+), Category: (.+)' that:
                  Captures the numeric score with (\d+) to match one or more digits
                  Captures the category name with (.+) to match the rest of the text on the line

            \item 2) This pattern will work correctly even if category names contain spaces or special characters;
        \end{itemize}

    \item 3. Data Extraction:
        \begin{itemize}
            \item 1) Using re.findall() to get all matches in the text
            \item 2) Each match will be a tuple with two elements: the score and category
        \end{itemize}
        
    \item 4. Data Processing: I iterate through each match to:
            \begin{itemize}
                \item 1) Convert the score to integer with int(match[0])
                \item 2) Keep the category as a string
                \item 3) Create a dictionary for each data point with proper column names
            \end{itemize}

    \item 5. DataFrame Creation:
            \begin{itemize}
                \item 1) Create a pandas DataFrame from the list of dictionaries
                \item 2) This automatically sets up the correct column names ("Score" and "Category")
                \item 3) The Score column is already converted to integers during the processing step
            \end{itemize}

\end{itemize}

~

\noindent
\textbf{Codes:}

\begin{lstlisting}[language=Python]
import pandas as pd
import re

def task_func(text):
    # Define the pattern to extract score and category
    pattern = r'Score: (\d+), Category: (.+)'
    
    # Find all matches in the text    
    matches = re.findall(pattern, text)

    # Create a list of dictionaries to store the data
    data = []
    for match in matches:
        score = int(match[0])  # Convert score to integer
        category = match[1]
        data.append({'Score': score, 'Category': category})

    # Create a DataFrame from the list of dictionaries
    df = pd.DataFrame(data)
    return df
\end{lstlisting}

\\
\bottomrule
\end{tabular}
\caption{\textbf{Example model outputs.} We pick a input from BigCodeBench where our model generates the correct answer. }
\label{fig:case}
\end{figure}

To provide qualitative insights into how \toolname approaches program repair tasks, we analyzed representative examples from BigCodeBench and SWE-bench Verified. These examples showcase the model's reasoning patterns and solution strategies for different types of programming challenges.  

\subsubsection{Structured Data Extraction}

In the BigCodeBench example depicted in Figure \ref{fig:case}, \toolname demonstrates a systematic approach to parsing structured text data. The reasoning trace reveals a clear problem decomposition strategy:

(1) Problem analysis and identification of key sub-tasks;
(2) Regex pattern design with explicit consideration of edge cases;
(3) Data extraction with appropriate type conversion;
(4) Structured DataFrame creation.

This reasoning pattern mirrors the approach a human programmer would take, breaking down the problem into manageable components and addressing each methodically. The implementation correctly handles score conversion to integers—a critical requirement—while maintaining a clean, readable solution.

\subsubsection{Type Coercion Bug Fix}
Due to space requirements, verbatim model output for this use case on Django from  SWE-bench Verified are reported in the supplementary file.

For the Django bug fix case, \toolname demonstrates deeper reasoning capabilities related to program repair:

\noindent
1. Precise bug identification (coercion of lists to tuples)

\noindent
2. Root cause analysis in the existing code

\noindent
3. Principled solution development (preserving input type with \texttt{type(value)})

\noindent
4. Verification considerations (test cases for different input types)

The generated patch is functionally equivalent to the ground truth solution, using \texttt{type(value)(...)} to dynamically reconstruct the output with the same type as the input. This demonstrates \toolname's ability to lead the base model toward understanding subtle programming concepts like type preservation and iterative data transformation.

\rqanswer{5}{
\toolname's reasoning traces reveal systematic problem decomposition, targeted debugging, and principled solution development—closely mirroring expert human problem-solving approaches.
}

\subsection{Generalization Capability}

To assess \toolname's ability to generalize across different code-related tasks, we conducted cross-benchmark experiments. We fine-tuned the model on SWE-bench Verified (500 data points) and evaluated it on the full BigCodeBench dataset (1,140 data points).

\toolname achieves 32.7\% Pass@1 on the full BigCodeBench dataset and 33.4\% on the test split. While this represents a slight performance drop compared to direct fine-tuning on BigCodeBench (35.6\%), it still substantially outperforms the base model (30.8\%), demonstrating effective transfer learning across different code-related tasks.


\rqanswer{6}{
\toolname demonstrates strong generalization capabilities, successfully transferring reasoning skills across different code-related tasks and benchmarks without requiring task-specific fine-tuning.
}

\subsection{Summary of Results}

Our experimental evaluation demonstrates that \toolname successfully closes the performance gap between open and closed-source models across diverse program repair and code generation benchmarks. The approach provides the most substantial improvements on complex software engineering tasks, with gains of up to 24.5\% in absolute performance. Both components of our methodology—Reasoning Trace learning and RL-LF—contribute significantly, with RL-LF proving particularly impactful.

The qualitative analysis reveals that \toolname develops systematic reasoning patterns similar to human experts, including structured problem decomposition, principled solution development, and verification considerations. Furthermore, the model demonstrates strong generalization capabilities, transferring reasoning skills across different code-related tasks.

These results validate our core postulate that transferring reasoning capabilities from closed-source to open-source models can substantially improve performance on complex software engineering tasks, enabling organizations to leverage high-quality program repair capabilities while maintaining the flexibility, customizability, and privacy advantages of open-source models.

\section{Discussion}
\label{sec:discussion}

This section examines the implications of our experimental results, discusses limitations, and explores future directions.

\subsection{Analysis of Key Findings}

Our experimental results highlight several important insights about reasoning transfer between LLMs for program repair tasks.

\subsubsection{Effectiveness of Reasoning Transfer}

The significant improvements achieved by \toolname, particularly on complex software engineering tasks, demonstrate that transferring reasoning processes—not just outputs—from closed-source to open-source models is highly effective. The varying impact across benchmarks suggests that explicit reasoning becomes increasingly valuable as task complexity increases, with the largest gains observed on SWE-bench Verified (+24.5\%). This aligns with research on the importance of reasoning in LLMs~\cite{wei2022chain, kojima2022large}.

\subsubsection{Synergy Between RT Learning and RL-LF}

The consistent pattern across all benchmarks shows that while Reasoning Trace (RT) learning provides modest but important improvements, RL-LF drives substantially larger gains. This complementary relationship suggests that RT learning enables the model to develop systematic problem-solving approaches, while RL-LF refines these capabilities based on implicit quality judgments that are difficult to capture through demonstrations alone.

The ablation studies reveal that this synergy is particularly powerful for complex tasks, where RT learning provides a foundation of structured reasoning that RL-LF can then optimize. This two-stage approach addresses limitations of previous methods that focused solely on either imitation learning or reinforcement learning.

\subsection{Limitations and Future Work}

While \toolname demonstrates significant progress toward bridging the gap between open and closed-source models, several limitations warrant consideration.

First, our approach depends on access to a high-quality closed-source model, creating a bootstrapping challenge. Future work could explore iterative improvement where each generation of enhanced open-source models helps train subsequent ones, gradually reducing this dependency.

Second, computational efficiency remains challenging due to the verbose nature of reasoning traces. More efficient methods for reasoning transfer could improve scalability, perhaps by identifying and focusing on the most informative parts of reasoning traces~\cite{li2023symbolic}.

Third, while \toolname shows strong generalization capabilities, maintaining 92\% of its performance when transferring from SWE-bench to BigCodeBench, this indicates some degree of benchmark-specific learning. Future work should explore techniques to enhance cross-domain generalization, possibly through meta-learning approaches or more diverse training data.

Future research directions could include:
\begin{itemize}
    \item Multi-model reasoning ensembles that combine insights from different models
    \item Task-specific reasoning strategies tailored to different types of programming problems
    \item Integration of human feedback to refine reasoning patterns before fine-tuning
\end{itemize}

\subsection{Implications for Software Engineering}

The performance of \toolname, approaching and sometimes exceeding closed-source models, has significant implications for software engineering practice.

By narrowing the gap between open and closed-source models, \toolname helps democratize access to advanced program repair capabilities. Organizations with privacy or customization requirements can now leverage high-quality alternatives without relying on API-based solutions.

Additionally, the explicit reasoning traces generated by \toolname could serve educational purposes, exposing novice programmers to structured problem-solving approaches. This transparency in reasoning also addresses a key limitation of current tools, where solutions are provided without explanation.

\subsection{Ethical Considerations}

Using closed-source models to improve open-source alternatives raises questions about intellectual property and appropriate attribution. While our approach does not extract model weights or architecture details, it does transfer knowledge embodied in outputs. Future work should explore frameworks for ethical knowledge transfer that balance innovation with appropriate attribution.

Responsible deployment should include integration with existing security tools, clear communication of model limitations, and ongoing monitoring for unintended consequences.

\section{Related Work}
\label{sec:related}
Our work builds upon and extends several research directions in program repair, knowledge transfer between language models, and reasoning enhancement.
\subsection{Program Repair with LLMs}
Recent work has demonstrated the effectiveness of LLMs for automated program repair tasks. ChatRepair~\cite{zhao2023chatrepair} and Repilot~\cite{wei2023copiloting} leverage ChatGPT's capabilities for bug fixing, while RING~\cite{dakhel2023github} focuses on integration with development workflows. Our approach differs by specifically addressing the performance gap between open and closed-source models through reasoning transfer, rather than developing methods that rely on proprietary models.
\subsection{Knowledge Transfer Between LMs}
Knowledge distillation techniques have been widely explored for transferring capabilities between models of different sizes~\cite{hinton2015distilling, sanh2019distilbert}. In the code domain, CodeDistill~\cite{wang2023codedistill} demonstrated distillation for code generation tasks. Our work extends these approaches by focusing specifically on transferring reasoning capabilities rather than general language abilities, and by combining distillation with reinforcement learning for optimal results.
\subsection{Reasoning Enhancement in LLMs}
Chain-of-thought prompting~\cite{wei2022chain} and similar techniques have shown the value of explicit reasoning for complex tasks. Program repair presents unique challenges for reasoning, as explored by ARCADE~\cite{jin2023program} and PACE~\cite{zhang2023planning}. Our contribution is the development of a systematic approach to extract, filter, and transfer reasoning patterns from closed-source to open-source models, combined with preference-based optimization.
\subsection{Reinforcement Learning for Model Alignment}
Reinforcement learning from human feedback (RLHF) has become a standard approach for aligning language models with human preferences~\cite{christiano2017deep, ouyang2022training}. Recent work has explored using AI feedback instead of human annotations~\cite{lee2023rlaif, bai2022constitutional}. Our RL-LF approach builds upon these foundations but specializes them for program repair by incorporating code-specific evaluation criteria and efficient preference data collection.
\section{Conclusion}
\label{sec:conclusion}
We presented \toolname, a methodology for enhancing open-source LLMs through reasoning transfer and reinforcement learning with LLM feedback. Our experimental results across four benchmarks demonstrate substantial improvements (up to +24.5\% on SWE-bench Verified), significantly narrowing the performance gap with closed-source models. Both components—reasoning trace learning and RL-LF—contribute to these gains, with their synergistic effect most pronounced on complex tasks requiring multi-step reasoning. This work shows that reasoning transfer is a viable approach for improving open-source models without accessing proprietary data or weights, enabling organizations to leverage high-quality program repair capabilities while maintaining the advantages of open-source deployment. Future work could extend this approach to other software engineering tasks and explore more efficient methods for reasoning transfer.

\noindent{\bf Open Science.}
\label{sec:data_availability}
To promote transparency and facilitate reproducibility, we make our artifacts available to the community at: 
\begin{center}
\url{https://anonymous.4open.science/r/REPARITY-9BE1}.
\end{center}
The repository includes the experiment scripts and the results as well as the checkpoints. In addition, A supplementary file containing complementary analysis has been included to provide further details about this work.

\bibliographystyle{ACM-Reference-Format}
\bibliography{refs}

\end{document}